%% file: a-main.tex
\documentclass[sigplan,10pt]{acmart}

\usepackage{booktabs} 
\usepackage[ruled,lined]{algorithm2e} 
\usepackage{amsmath}

\begin{document}

\title{i13DR: A Real-Time Demand Response Infrastructure for Integrating Renewable Energy Resources}

\titlenote{This is the comprehensive version of the work published at e-Energy '18: Proceedings of the Ninth International Conference on Future Energy Systems, DOI: https://doi.org/10.1145/3208903.3212053~\cite{idr}.}

\author{Pezhman Nasirifard}
\affiliation{
  \institution{Technical University of Munich}
   \country{Germany}
}
\email{p.nasirifard@tum.de}

\author{Hans-Arno Jacobsen}
\affiliation{
  \institution{University of Toronto}
  \country{Canada}
}

\renewcommand{\shortauthors}{P. Nasirifard, et al.}

\begin{abstract}
With the ongoing integration of Renewable Energy Sources (RES), the complexity of power grids is increasing. Due to the fluctuating nature of RES, ensuring the reliability of power grids can be challenging. One possible approach for addressing these challenges is Demand Response (DR) which is described as matching the demand for electrical energy according to the changes and the availability of supply. However, implementing a DR system to monitor and control a broad set of electrical appliances in real-time introduces several new complications, including ensuring the reliability and financial feasibility of the system. In this work, we address these issues by designing and implementing a distributed real-time DR infrastructure for laptops, which estimates and controls the power consumption of a network of connected laptops in response to the fast, irregular changes of RES. Furthermore, since our approach is entirely software-based, we dramatically reduce the initial costs of the demand side participants. The result of our field experiments confirms that our system successfully schedules and executes rapid and effective DR events. However, the accuracy of the estimated power consumption of all participating laptops is relatively low, directly caused by our software-based approach.
\end{abstract}

\keywords{Smart Grid, Demand Response, Demand Side Management (DSM), Load Management}

\settopmatter{printacmref=false}
\setcopyright{none}
\renewcommand\footnotetextcopyrightpermission[1]{}
\pagestyle{plain}
\settopmatter{printfolios=true}

\maketitle


\input{sections/introduction}
\input{sections/related_work}
\input{sections/approach}
\input{sections/evaluation}

\input{sections/discussion}
\input{sections/conclusions}

\begin{acks}
We sincerely thank Jose Rivera and Martin Jergler for their valuable input to this work and for advising on the master's thesis on which this paper is built~\cite{pezhman_thesis}. 
\end{acks}

\bibliographystyle{ACM-Reference-Format}
\bibliography{bibliography} 

\input{sections/appendix}

\end{document}

%% file: sections/introduction.tex
\section{Introduction} 
\label{sec:introduction}

Reduction of greenhouse gasses is one of the significant concerns of the global community, where the continuous development of Renewable Energy Sources (RES) plays a crucial role \cite{ParisAgreement}. Despite several apparent advantages, integrating RES into existing transmission and distribution grids is challenging. One main issue is the fluctuating nature of RES, which increases the complexity and vibrancy of the power grids \cite{cost_efficient}. Several works address these issues by proposing a broad range of solutions, ranging from developing more energy-efficient appliances and materials to developing Smart Grids (SG) that actively match the demand side's load with the available energy supply. This concept is known as Demand Response (DR) \cite{survey_dr_1, survey_dr_2, demand_response_advanced_metering}. 

A minimal DR infrastructure consists of a few essential components, including a power measurement and control component for estimating and managing the load on the demand side, an overlay network for connecting the demand-side participants with the electrical utilities who aggregate the power consumptions and schedule DR events to match the demand with the available supply. We define the DR event as the specific schedule during which the appliance's power consumption must be managed to match the available supply. Participation of many electric appliances on the demand side plays a crucial role in implementing a robust DR infrastructure. However, managing many distributed devices requires a sophisticated and resilient system. This complexity increases with the integration of immediate fluctuating RES. Another challenge is providing the demand-side participants with attractive incentives to join the DR system. The participants are required to initially pay for enabling their appliances with power measurement and control devices and communication systems. This initial cost discourages several potential participants, mainly residential consumers, who might not receive a significant financial gain from participating in the DR events \cite{demand_response_in_samrt_electricity}. 

In this work, we propose a design and implementation of a distributed DS infrastructure for balancing the power consumption of a network of laptops with the intermittent supplies of RES in real time. The primary reason for choosing laptops as our target appliances is the extensive prevalence of laptops in everyday use. Furthermore, we use laptops' computational resources and communicational capabilities to offer an entirely software-oriented approach for performing DR tasks and events, including monitoring, estimating, and controlling the power consumption of laptops and interacting with the utilities. Furthermore, we design and conduct field experiments with some participating laptops and a simulated RES to evaluate the performance of our design for performing DR events.

Several DR infrastructures have already been realized and implemented \cite{day_1, day_2, day_3}, but they require a significant time ahead for scheduling and performing DR events. Our approach's novelty is developing a responsive DR system where DR schedules are created and executed immediately upon a request, e.g., when the production of RES suddenly drops. Furthermore, since our approach is entirely software based on the demand side, we dramatically reduce the initial cost of demand-side participants to zero cents.

We recognize the following contributions to this work:

\begin{enumerate}
\item We offer a design and implementation of a responsive DR system for matching the power consumption of several connected laptops with the fluctuating supply of RES. Our systems schedule and creates DR events immediately when changes in RES occur and notify the participating devices in real time. 
\item To increase the motivation of the demand-side participants to join the DR system, we offer an entirely software-based approach, which significantly reduces the initial cost to zero. Our demand-side approach is capable of measuring laptops' current power consumption using regression models. Our system also controls the power consumption of laptops by using the built-in functionalities of operating systems (OS). 
\item The majority of the academic research on the different aspects of DR is based on mathematical simulations without any empirical measurements of real-world experiments \cite{smart_in_home_power_scheduling, optimal_power_scheduling_method}. We offer a DR system that fills this gap and enables the researchers with a testbed for experimenting with different DR optimization and scheduling approaches.
\end{enumerate}

We organize the rest of the paper as follows: First, in Section \ref{sec:relatedWork}, we review the existing literature on DR infrastructures and power modeling. Then, we analyze the realized requirements of our system and explain our implementation in Section \ref{sec:approach}. Afterward, in Section \ref{sec:evaluation} we explain the conducted experiments to evaluate the performance of our system. We discuss our findings and the limitations of our approach in Section \ref{sec:disc}. Finally, in Section \ref{sec:conclusion}, we summarize our work and discuss future works.

%% file: sections/related_work.tex
\section{Related Work} 
\label{sec:relatedWork}

RES accounts for 28\% of electricity production in Germany, and it has been steadily increasing over the past two decades \cite{german_elec_usage}. Nevertheless, the fluctuating nature of RES, commonly the wind and the sun, causes many concerns with the irregularity, uncertainty, and unreliability of supplies \cite{cost_efficient, ogm_drone, ogm_energy}. One solution for increasing the predictability of the RES is a weather forecast which is not without errors \cite{wind_forecast}. The typical solution is using on-site or emergency generators, which are expensive, polluting, and face many governmental restrictive regulations in many countries \cite{survey_dr_1}. Also, since RES is often connected to the power grid at the distribution level, e.g., in Germany, 90\% of installed RES is directly connected t the distribution grid, the unpredictability of the power consumption on the demand-side increases the complexity of the system. Several studies confirm DR as a useful solution for addressing these issues \cite{combined_operations_of_renewable_energy, dr_res_not_impel_1, dr_res_not_impel_2, dr_res_not_impel_3}. In \cite{combined_operations_of_renewable_energy}, authors offer an approach for optimizing the integration of RES to the grid by using energy management systems and offering real-time pricing, where the simulation study yields good results for matching the demand with available RES. In \cite{dr_res_not_impel_2, dr_res_not_impel_3}, the authors offer scheduling and optimization approaches for maximizing the benefits of DR in the presence of volatile RES and reducing the electricity cost. Their simulation study also confirms DR as a good candidate for integrating RES into the grid. We extend the previous works by implementing a real-world DR infrastructure where different optimization and scheduling can be realized through field experiments. Moreover, we provide a real-world evaluation of the approach, which was missing in the literature.


Residential and small commercial consumers account for about 40 percent of total electricity consumption in Germany \cite{german_elec_usage}, and their share will increase with the adaptation of electric vehicles~\cite{felix}. They can contribute to a DR system by curtailing or shifting their demand loads to create a more reliable distribution and transmission grid~\cite{ogm_energy, ogm_johannes}. Reducing the initial cost of joining the DR system is a significant incentive for residential and commercial consumers. Therefore, we use the built-in functionalities of consumers' laptops to reduce the initial cost. We achieve this by measuring the power consumption of laptops by using mathematical power models. Several studies have explored the power consumption models for server environments and data centers \cite{data_center_energy_consumption, comparison_high_level_full_system, no_hardware_required, thomas_thesis}. However, we can not directly apply these works to laptops because server environments and laptops consume energy differently. Studies on constructing power models for laptops and battery-operated electrical appliances are relatively limited \cite{self_constructuve_high_rate_system_modeling}, mainly because of the laptops' limited consumption in comparison to server environments. According to \cite{comparison_high_level_full_system}, an appropriate mathematical model for estimating power consumption must satisfy several requirements. These constraints include the accuracy of the power estimation, the speed of prediction, the generality of the model when applied to various systems with different hardware and software specifications, affordable, non-intrusive measuring devices, and the simplicity of the design. Furthermore, \cite{comparison_high_level_full_system} classifies real-time power modeling approaches into two groups: detailed analytical power modeling and high-level black-box modeling. The analytical power modeling exploits the CPU performance counters for accurately estimating the power. However, this approach only applies to particular processors that are not portable from one system to another. The high-level black-box modeling technique uses system metrics, such as CPU, disk, and memory utilization, for constructing linear or multiple regression power estimation models. This type of modeling is less accurate than analytical modeling; however, black-box modeling is more general and portable due to its independence from system specifications. Furthermore, \cite{comparison_high_level_full_system} constructs and evaluates five different power consumption models by using \textit{Mantis} \cite{full_system_power_analysis}, a non-intrusive power modeling system. Mantis uses a one-time model fitting, during which system utilization metrics are fitted to power readings of an external AC power meter. Evaluations of the generated power models indicate that the utilization-based regression models perform better than other models. We should mention that the discussed studies fit and generate power models using external assistance like AC power meters or a second computer.
In contrast, \cite{self_constructuve_high_rate_system_modeling} introduces an approach for self-constructing power modeling systems for laptops and mobile devices. This approach exploits the smart battery interface of laptops for self-power measurement. Despite the accurately estimated power consumption, the constructed power models only measure the power consumption of laptops without considering the AC battery charger. In contrast, measuring the power consumed by laptops' AC chargers is essential for our work. In this study, we follow the same approach for collecting system utilization metrics and actual power consumption of laptops by using external power meters for creating linear regression power consumption models for laptops. However, we deploy the generated models on other laptops without refitting the models to examine the model's portability. 

We differentiate ourselves from previous works in integrating the RES with a real-time responsive DR system and with non-intrusive power consumption modeling, i.e., we only estimate the power consumption by measuring the laptop utilization metrics. Finally, in contrast to other works, we conduct field experiments to evaluate the performance of implemented DR system and power models. 

%% file: sections/approach.tex
\section{Design and Implementation of 13DR} 
\label{sec:approach}

In the following section, we discuss the realized use cases and explain the implementation of our DR infrastructure in detail. For the rest of the paper, we call our proposed DR system \textit{i13DR}, where i13 refers to the location where this study has been conducted, and  DR stands for Demand Response. We open-sourced the code~\footnote{\url{https://github.com/i13DR}}

\subsection{Overview} 
\label{sec:app_overview}

\begin{figure}[t!]
\centering
\includegraphics[width=1\columnwidth]{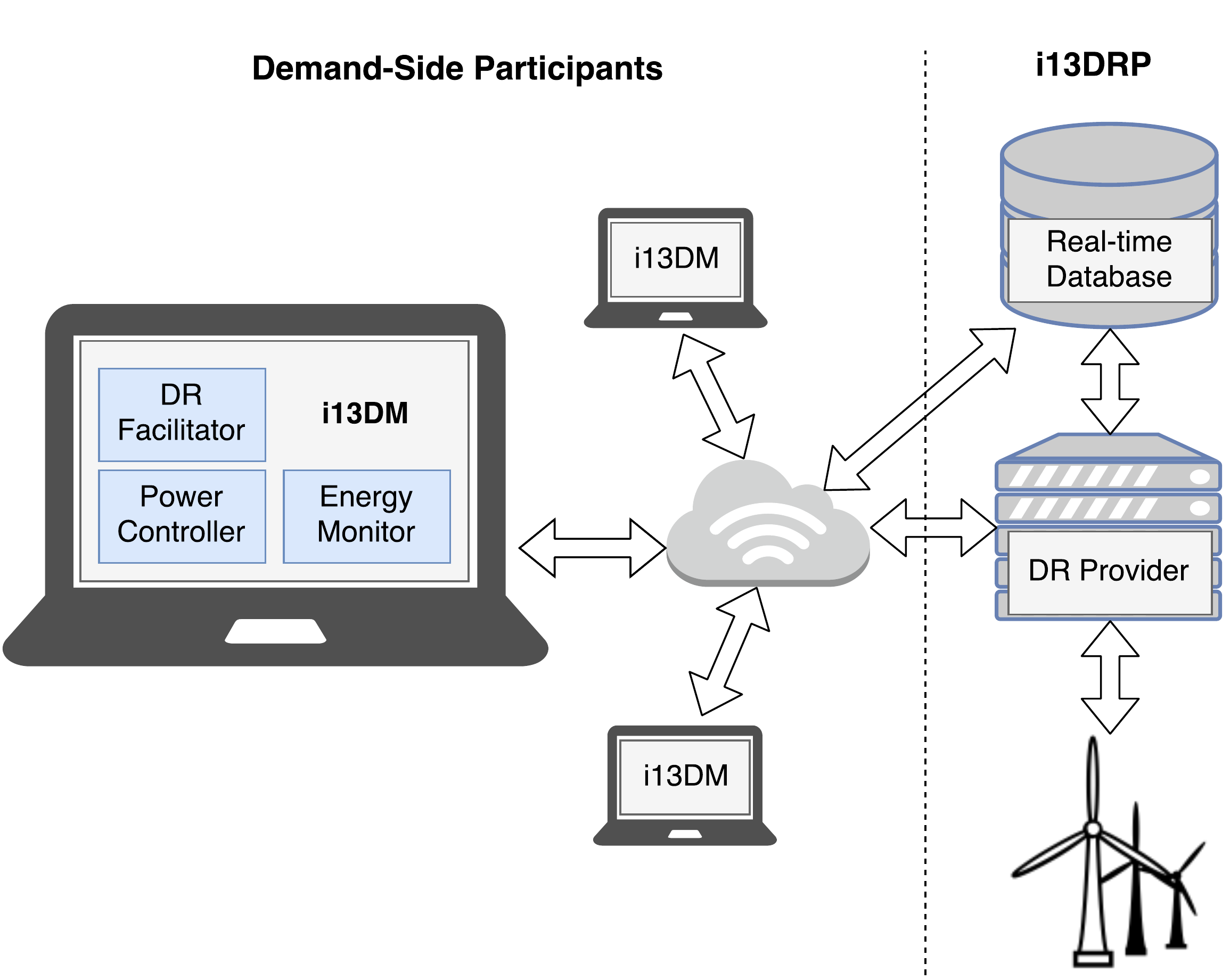}
\caption{Overview of i13DR infrastructure design.}
\label{fig:designi13DR}
\end{figure}

The proposed i13DR design consists of two major parts, a demand-side manager application and a bundle of server-side applications that manage the DR system, as Figure \ref{fig:designi13DR} displays. For the rest of the paper, we refer to the demand-side application as \textit{i13DM}, which stands for \textit{i13 Demand Manager} and we call the server-side DR provider \textit{i13DRP}, that stands for \textit{i13 Demand Response Provider}. We design and develop the i13DM to encapsulate the fundamental functionalities for performing an efficient DR event, including the features for measuring and limiting the power consumption of laptops and communicating with the i13DRP. On the server side, the i13DRP consists of two primary subsystems, the DR provider and the real-time database. The DR provider is responsible for managing the laptops and communicating with RES to inquire about the availability of supplies. Furthermore, i13DRP performs the scheduling and execution of the DR events with the cooperation of a real-time database. The real-time database persists the DR-related data and can distribute the data in real time. 

\subsection{Requirements Analysis of i13DR} 
\label{sec:app_usecases}

\begin{figure*}
\includegraphics[width=1.7\columnwidth]{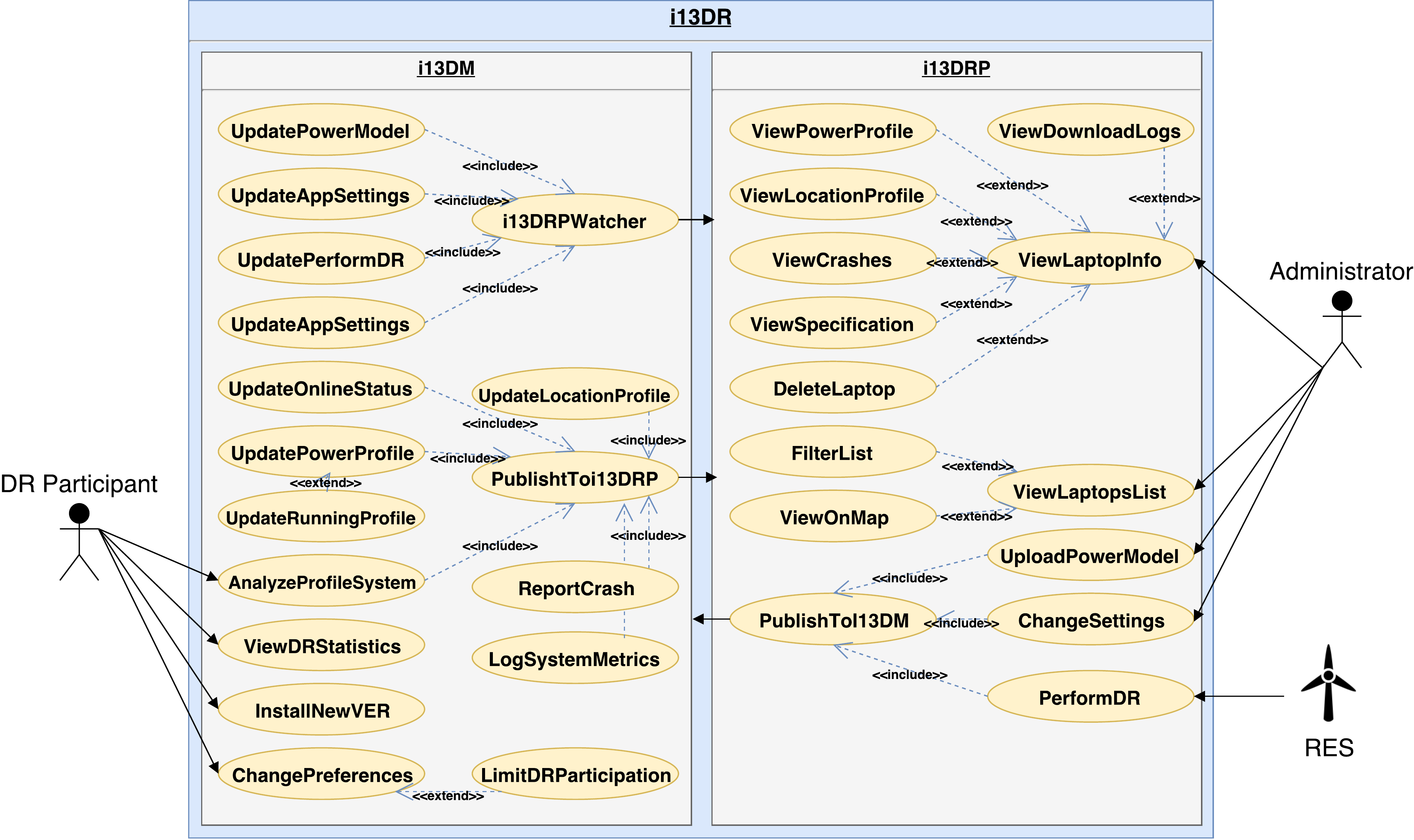}
\caption{Use case diagram of i13DR.}
\label{fig:usecasedisgram}
\end{figure*}

We realize several use cases for i13DR, and to better understand the scope of i13DR, we utilize a use case diagram to summarize the use cases as Figure \ref{fig:usecasedisgram} displays. A use case describes the system's functionality that the user comprehends \cite{oop_book}. Typically an actor is described as an entity interacting with the system, including a user, another system, or the system's physical environment. However, to enhance the understandability of our diagram, since i13DR internally consists of two parts of i13DM and i13DRP interacting with one another, we recognize both i13DM and i13DRP as the studied systems as well as actors. Figure \ref{fig:usecasedisgram} shows that we have three external actors. On the left, we have the DR participant who directly interacts with i13DM. On the right side of the diagram is the administrator of i13DRP and the RES interacting with i13DRP, and finally, in the middle, i13DRP and i13DM communicate with each other.

For performing the DR-related activities, i13DR relies on two primary datasets of \textit{Location Profiles} and \textit{Power Profiles} of laptops which the i13DMs provide and continuously update. Location profile provides the i13DRP with an approximate location of the laptop during the week, i.e., i13DRP, with some probability, knows about the laptop's location at the specific time of the week. The power profile of each i13DM provides an estimate of laptops' power consumption in standard mode and power save mode during a week. We define the power save mode as the time when the i13DM activates the load control mechanism on the laptop to reduce the power consumption, which we achieve by taking advantage of the power saving functionalities of OS. We require managing the profiles because we assume people follow a weekly routine, e.g., the laptops are located five days a week at the users' works and two days at the users' homes. Thereby, we can create context-aware DR events depending on the requirements of the grid in different locations.  

In the following, we describe the five most important use cases in detail:

\subsection{Perform DR Event} 

This use case explains the steps for executing a successful DR event in the i13DR system. The i13DRP continuously monitors the current production of the RES. When i13DRP detects a reduction of supply, i13DRP selects the candidate laptops according to existing power profiles and location profiles of laptops connected to the same power grid as RES. Then i13DRP uses the scheduling algorithm (any optimization and scheduling algorithm can be plugged and play, we do not implement a specific scheduling approach for this study) to estimate the power reduction that candidate laptops can offer in a way that the estimated reduction matches the current supply. Once i13DRP decides on the participating laptops, i13DRP creates a DR schedule for the event containing the start time and duration of the DR event and submits the schedule to the real-time database. Finally, i13DM on the candidate laptops is immediately notified about the new schedule through the real-time database. The i13DMs fetch the schedule from i13DRP, update their local schedule accordingly, and activate the power control at the appropriate time.

\subsection{Initial Profiling and Analysis of Laptop} 

This use case only executes once when i13DM is launching for the first time on the user's laptop to collect the necessary data about the laptop's specifications and create the initial location and power profiles. Upon the initial startup, i13DM executes several platform-specific commands to extract the hardware and software specifications. Once the analysis is over, i13DM submits the extracted data to i13DRP. This data is later used for estimating power consumption. Afterward, i13DM fetches the previously constructed generic power model from i13DRP for calculating the power consumption of the laptop in normal mode and power save mode. Then, i13DM creates a power profile dataset containing an object for every minute of the day and every day of the week. After, for every object in the dataset, i13DM estimates the normal mode and power save mode power consumptions calculated based on the power models. Also, i13DM fixes the probability of the laptop running, the i13DM application running, and the laptop being connected to the AC adapter to 50\%. After, 13DM stores the power profile in the local database and submits the profile to the i13DRP. After, i13DM uses geolocation APIs to locate the current position of the laptop. Then, i13DM creates a location profile dataset containing an object for every quarter of an hour and every day of the week. Afterward, for every object in the dataset, i13DM sets the longitude, latitude, measurement accuracy, and the location's zip code. Furthermore, i13DM fixes the probability of the laptop being presented in this area to 100\%. Finally, once the location profiling is over, i13DM stores the location profile in the local database and submits the location profile to i13DRP. After finishing the initial system analysis and profiling, the i13DM  keeps running continuously in the background. 

\subsection{Update Location Profile} 

i13DM regularly updates the location profile of the laptop. Every 90 seconds, i13DM locates the current geographical position of the laptop by using geolocation API and inserts the longitude, latitude, accuracy of measurements, and zip code into its local database. After, i13DM queries the local database to retrieve the locations with the time of the day and the weekday equal to the last location record inserted. Then, i13DM groups the retrieved records by zip code and selects the longitude and latitude of the group with the highest number of records and sets the probability of the laptop being present in this longitude and latitude by dividing the number of records in the group by the number of all fetched records. Finally, i13DM updates the corresponding location profile in the local database and submits the new location profile to i13DPR.

\subsection{Update Power Profile} 

i13DM continuously updates the power consumption profile of the laptops. Every few seconds (one-second interval on Ubuntu and three seconds interval on Windows machines. We use different intervals due to some platform-related limitations), i13DM uses the power model to estimate the current power consumption of the laptop in normal mode and power save mode and checks if the laptop is connected to the AC charger. Then, i13DM inserts the record into the local database. After, i13DM queries the local database to retrieve the stored power readings, which have the time of the day and the weekday equal to the last observed power reading, and groups the records based on the status of being connected to the AC charger. After, i13DM calculates the arithmetical mean of power consumption in normal mode and power save mode for all retrieved records in each group and sets the probability of the laptop being connected to the AC charger by dividing the number of records in each group by the total number of fetched rows. Finally, i13DM updates the corresponding power profile record in the local database with new measurements and submits the updated power profile to i13DRP.

\subsection{Update Running Profile} 

Every 90 seconds, i13DM inserts an \textit{activity record} to the database, indicating that i13DM and laptop are running. Afterward, i13DM retrieves the two most recent activity records and calculates the time difference between the two retrieved records. If the time difference is more significant than 90 seconds, then i13DM inserts several new activity records into the database containing a status flag indicating that i13DM and the laptop were not running within the calculated time difference. After, i13DM queries and groups the stored activity record for every minute of the day and every day of the week. Then, for every group, i13DM calculates the probability of i13DM and laptop running by dividing the number of records with the status flag by the total number of rows in that group. Finally, i13DM updates the corresponding power consumption profiles at the same time of the day and the day of the week in the local database and submits the updated power profile to i13DRP. \\

While designing the i13DM, we ensure that i13DM performs DR-related tasks independently without any required interaction with the user. Although i13DM does not interrupt the user's activity, we provide the user with the option of preventing any DR activity when desired. We also have to emphasize that we only collect the laptops and users' data that we specifically require for performing DR activities. We sufficiently anonymize the collected data and transfer the data to i13DRP through secure and encrypted connections. Although we designed the i13DRP so that DR-related activities can be performed through direct interaction with utilities and RES without the supervision of any administrators, we provide a control panel for the administrator to manage the registered laptops, their data, and the DR-related tasks. 

\begin{figure*}
\includegraphics[width=1.5\columnwidth]{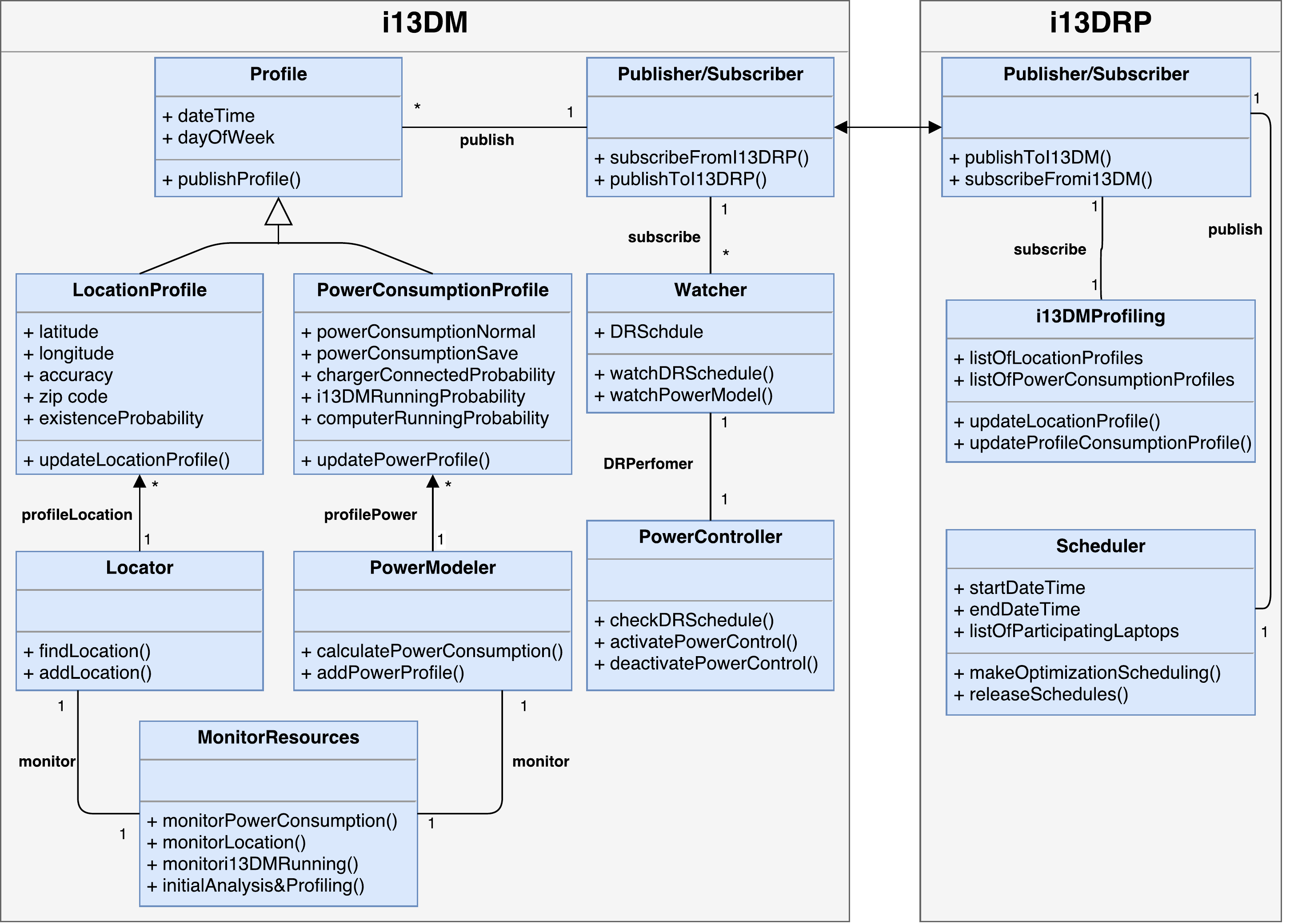}
\caption{Class diagram of i13DR.}
\label{fig:classdiagram}
\end{figure*}

We describe the structure of i13DR by using a class diagram based on the discussed use cases and requirements, as Figure \ref{fig:classdiagram} displays. Classes are abstractions of the behaviors and attributes of the system, and objects are entities that encapsulate behavior and state of components \cite{oop_book}. On the left side of Figure \ref{fig:classdiagram}, we present the i13DM classes, and on the right side, we include the classes for i13DRP. The i13DM classes characterize the system required for power and location profiling and performing DR events. \textit{MonitorResources} in association with \textit{Locator} and \textit{PowerModeler} initiates the activities for profiling and monitoring resources. Furthermore, \textit{Watcher,} subscribes to \textit{Publisher/Subscriber} for receiving power models and DR schedules, and \textit{PowerController} is responsible for activating and deactivating the power controls according to the schedules. On the i13DRP side, \textit{i13DMProfiling} subscribes to \textit{Publisher/Subscriber} for obtaining the power consumption and location profile from i13DM and \textit{Scheduler} is responsible for scheduling a DR events.

\subsection{Implementation of i13DR} 
\label{sec:impel}

\begin{figure*}
\includegraphics[width=1.5\columnwidth]{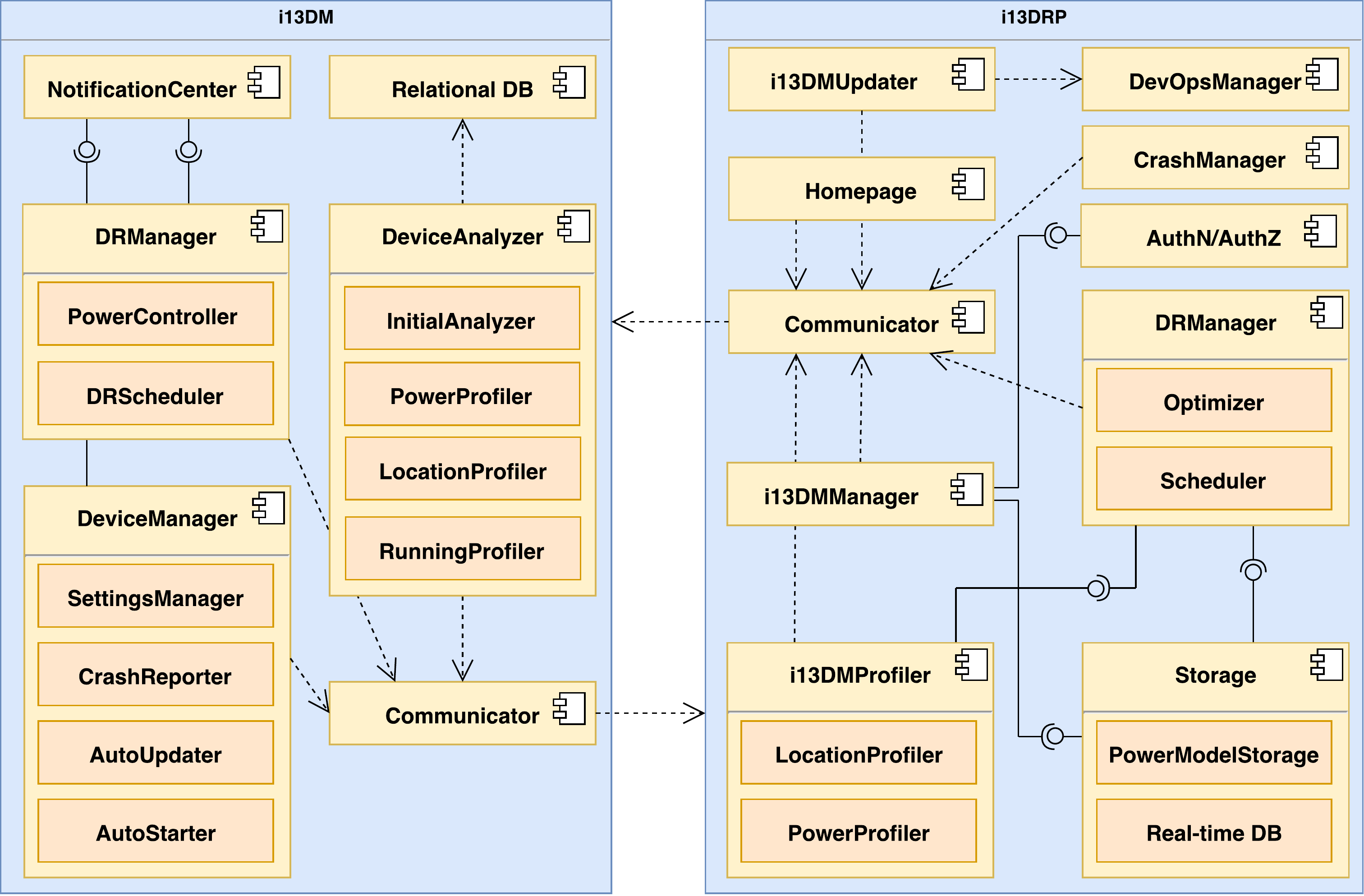}
\caption{Subsystem decomposition of i13DR.}
\label{fig:decomposition}
\end{figure*}

Before implementing the i13DR, to decrease the complexity of our design, we break down the i13DM into smaller resizeable subsystems according to the recognized use cases. We decompose the i13DR into subsystems with low coupling and high coherence to decrease the dependencies between two or multiple subsystems and increase the dependencies among classes within a subsystem, as Figure \ref{fig:decomposition} illustrates. \textit{DRManager}, \textit{DeviceManager} and \textit{DeviceAnalyzer} are the fundamental parts of i13DM carrying out the necessary tasks of a DR event. \textit{DeviceAnalyzer} contains the entities responsible for estimating power consumption, location profiling, and initial analysis. \textit{DeviceManager} consists of classes responsible for the general behavior of the application. \textit{DRManager} holds all the classes for managing a DR event. 

As previously stated, i13DRP consists of several server-side applications, but we do not present the boundaries to keep the subsystem decomposition diagram clear. The primary task of i13DRP is managing the participating laptops and performing DR events. \textit{DRManager} carries out the essential scheduling operations in association with \textit{i13DMProfiler}, which is responsible for aggregating the power consumption and location profiles of laptops. \textit{i13DMManager} provides the administrator of i13DRP with the necessary operations for managing the registered laptops. Furthermore, several other subsystems facilitate i13DRP operations including \textit{CrashManager} handling the reported i13DM crashes, \textit{i13DMUpdater} and \textit{DevOpsManager} maintain building and releasing new versions of i13DM and i13DRP and finally, \textit{AuthN/AuthZ} maintains the authentication/authorization and security policies of i13DM and i13DRP. Furthermore, in the appendix, we include a few snapshots of the implemented i13DM and i13DRP.

\subsubsection{Implementation of i13DM} 
\label{i13DMImplementation}

We develop and package the discussed functionalities of i13DM into an cross-platform desktop application for Microsoft Windows 7/8/10 and Ubuntu 16.04 LTS based on \textit{Electron Framework}\footnote{\url{https://electron.atom.io/}} (a JavaScript framework for creating desktop applications) and several \textit{Node.js}\footnote{\url{https://nodejs.org/en/}} packges.

To reduce laptops' power consumption, we use the built-in power management features of OS. The ideal would be disconnecting the AC battery charger from the plug and only draining power from the battery. However, the OS does not provide any functionalities for this purpose. Hence, we use the OS API to put the laptop into power save mode. On Windows, we rely on \textit{powercfg}\footnote{\url{https://msdn.microsoft.com/en-us//library/hh824902.aspx}} to import, activate and deactivate Windows-specific energy save scheme, which we previously generated and exported from a Windows 10 machine and we ship it with i13DM. Ubuntu is equipped with very few built-in functionalities for controlling power consumption. For this reason, to save power, we only dim the screen and turn the screen off when the laptop is idle. 

To perform the initial analysis of laptops, we use built-in functionalities of OS to extract the hardware and software specifications. On Windows, we heavily make use of  \textit{Windows Management Instrumentation (WMI)}\footnote{\url{https://msdn.microsoft.com/en-us/library/aa384642(v=vs.85).aspx}} commands to query the system specification. However, on Ubuntu we rely on several commands, including but not limited to \textit{lshw}, \textit{dmidecode}, \textit{lshw}, \textit{lspci} and \textit{lscpu} to collect the necessary system specifications.

\subsubsection{Implementation of i13DRP} 
\label{i13DRPImplementation}

i13DRP comprises several server applications which emphasize real-time communication and scalability. We implement the primary functionalities of i13DRP based on \textit{Meteor}\footnote{\url{https://www.meteor.com/}} and \textit{Angular 2+}\footnote{\url{https://angular.io/}}, which are JavaScript frameworks for developing real-time back-end and front-end server applications. We also use \textit{NGIX}\footnote{\url{https://nginx.org/}}, a high-performance HTTP and reverse proxy server running on a Ubuntu 16.04 LTS to server the applications. 

The main storage provider of i13DRP is \textit{Firebase}\footnote{\url{https://firebase.google.com/}} providing us with a real-time cloud-hosted NoSQL database for storing and retrieving the i13DM and i13DRP data in JSON format. Firebase also offers cloud storage, which we use to distribute the constructed power models to i13DM in real time. The main reason for choosing this platform is the high scalability of the offered services, which are empowered by Google infrastructure to support several thousand simultaneous connections, reasonable prices, high security, and availability. 

\subsection{Power Model Construction} 
\label{sec:model_build}

\begin{figure}[t!]
\centering
\includegraphics[width=.6\columnwidth]{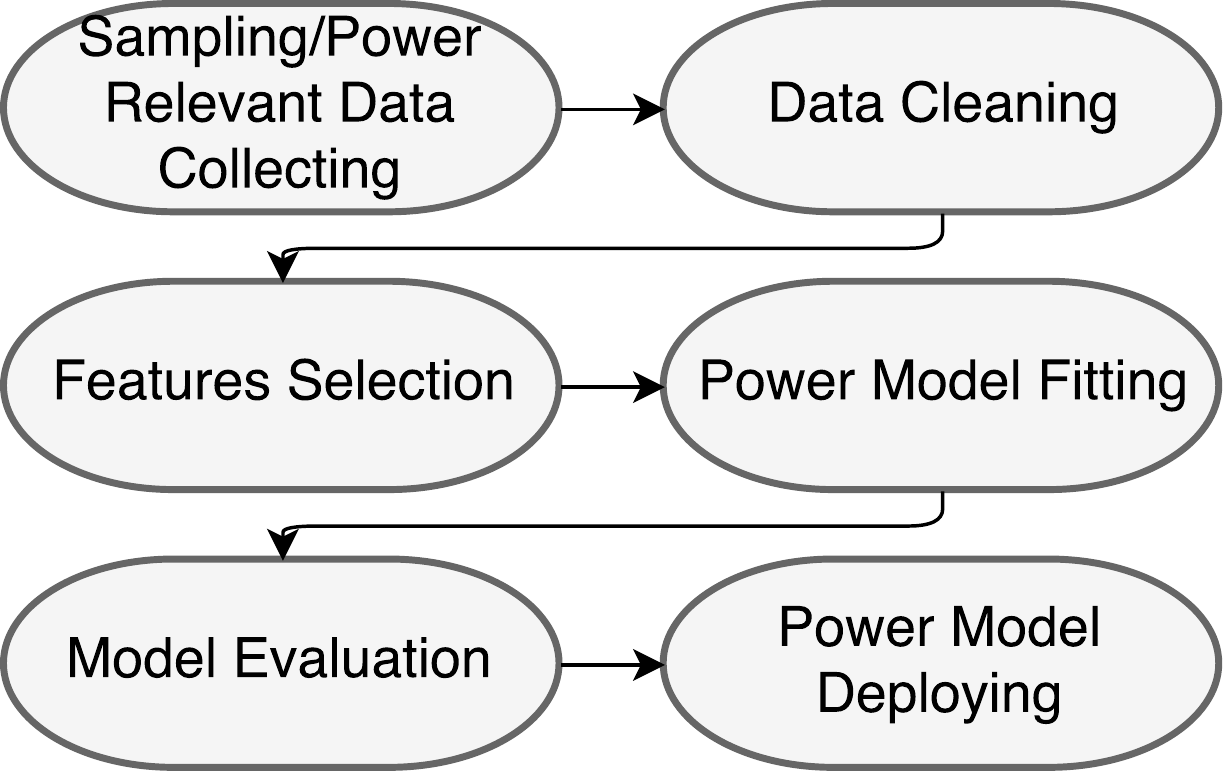}
\caption{Power model generation procedure.}
\label{fig:powermodelflow}
\end{figure}

We propose an approach for generating regression models, which i13DM uses to estimate the power consumption of laptops in real time based on the system metrics reading. The laptop's power consumption is positively correlated with the circuit's load. In order words, when the amount of power consumed is either load on the processors, networking traffic, or the screen's brightness increases. Also, while the laptop's battery is charging, the AC charger causes an extra load on the regular power consumption. Therefore, when estimating the power consumption, we extract the system metrics which highly correlate with the power consumption. Afterward, we feed the derived parameters into a previously built regression model to calculate the total power consumption. Equation \ref{eq:energyModel} shows the mathematical relationship between estimated power consumption \textit{$E_{estimated}$} and the power model \textit{$f()$}, where \textit{$m_1(t)$} to \textit{$m_n(t)$} are the extracted system metrics at the time \textit{$t$}. 

\begin{equation} \label{eq:energyModel}
E_{estimated} = f(m_1(t), ....m_n(t))
\end{equation}
Since i13DM is executable on Windows and Ubuntu and both OS have different specifications, we require creating four different power models: A model for Windows in normal mode, a model for Windows in power save mode, a model for Ubuntu in normal mode, and a model for Ubuntu in power save mode. Our modeling procedure is based on six distinct steps, as Figure \ref{fig:powermodelflow} displays.

First, we require collecting power consumption data for fitting the models. We use a \textit{Lenovo ThinkPad L540} with specifications described in Table \ref{table:540_specs} as SUT. We connect the SUT to \textit{MEDAL} \cite{medal}, a custom-built low-cost measurement system for high-frequency energy data built upon a  voltage-sensing circuit, current sensors, and a single-board PC as a data aggregator. Medal collects the active power consumption of SUT in watts with a sampling rate of up to 50 kHz. Meanwhile, we log the system metrics of SUT with a one-second interval on Ubuntu and a three-second interval on Windows. We use different intervals on different OS because of some platform-specific limitation that prevents us from faster reading. The logged system metrics include CPU Utilization in percentage, display brightness in percentage, power drain rate of battery in watts, the charging/discharging status, the remaining capacity of the battery in percentage, memory usage in percentage, disk read/write KB per seconds and request and network download/upload rate in KB per seconds. We collect the data over four days for each OS and power consumption mode while the user works with laptops performing the common daily activities. 

\begin{table}
\small
\begin{center}
\caption{Lenovo ThinkPad L540 Specifications}
\label{table:540_specs}

  \begin{tabular}{ ll}	
	\hline \hline	
	Vendor &  Lenovo\\
	Version & ThinkPad L540 (20AV002YGE) \\
	CPU & 	Intel - Core i5-4200M CPU @ 2.50GHz \\
	Video Card & Intel - 4th Gen Integrated Graphics \\
	Memory & 8GB SODIMM DDR3 \\
	Wireless interface &  Intel Wireless 7260 \\
	Storage & Seagate - ATA HGST HTS725050A7  \\
	Battery &  Sanyo - 45N1769 -56160mWh\\
	Operating System & Ubuntu 16.04 LTS / MS Windows 10 \\
\hline \hline
  \end{tabular}
\end{center}
\end{table}

After we preprocess, normalize and clean the data and remove the outliers from the raw collected data. We consider the power readings below 8 watts and above 65 watts (the maximum output of the used AC charger of the SUT) to be outliers and remove them from the dataset. Before training the linear regression model, to improve the accuracy and reduce the complexity of the constructed model, we examine the quality of different combinations of features to select a subset of features with the best accuracy. For this reason, we perform an exhaustive search to determine the best subset of power-related features. We include the detailed results of the exhaustive search of all models in the appendix. We perform the calculations by making use of \textit{R Programming Language}\footnote{\url{https://www.r-project.org/}}. According to the results, for the Windows in normal and power save mode, we select the following features:

\begin{enumerate}
\item Battery charging/discharging rate
\item Battery charging/discharging rate squared
\item Interaction of battery charging/discharging rate with display brightness
\item Interaction of battery charging/discharging rate with the remaining capacity of the battery
\item Charging status
\item CPU usage
\item Memory usage
\item Remaining capacity of the battery
\item Download/upload rate in KB
\item Disk read/write request per second
\end{enumerate}

We also select the following features for Ubuntu in normal and power save modes: 

\begin{enumerate}
\item Battery charging/discharging rate
\item Battery charging/discharging rate squared
\item Interaction of battery charging/discharging rate with the remaining capacity of the battery
\item Charging status
\item CPU usage
\item Memory usage
\item Remaining capacity of the battery
\item Download/upload rate in KB
\item Disk read/write request per second
\end{enumerate}

Before training the regression model with four datasets, first, we split each dataset into two smaller subgroups. One subgroup holds 80\% of the dataset that we use for fitting the model, and the other contains 20\% of the dataset meant for testing the model's accuracy in predicting out-of-sample values, which we discuss in the next section. Then, we use the selected features to fit a simple linear regression model by using the \textit{lm} method in \textit{R}. We should mention that we use the same power model generated on the SUT on different participating laptops.

%% file: sections/evaluation.tex
\section{Evaluation} 
\label{sec:evaluation}

In the following section, we evaluate the performance and accuracy of the i13DR and the constructed power models. First, we explain the methodologies we use for evaluating power models, followed by an experiment that we designed and conducted for quantifying the abilities of i13DR infrastructure. Then, we describe the objectives and hypotheses we expect to realize, followed by presenting the results gathered from the experiments. Afterward, we interpret and discuss the findings. Finally, we mention the limitation of our system. 

\subsection{Power Models Evaluation} 
\label{sec:powerModelsEvaluation}

We evaluated the four generated power models by using the test dataset of the corresponding model based on different metrics including \textit{Adjusted R-Squared}, \textit{MAPE}, and \textit{Min/Max Accuracy} of constructed models and also the \textit{Correlation Accuracy} between the actual and predicted values, as Table \ref{table:powerModelEvaluation} summarizes. The Adjusted R-Squared\footnote{\url{http://r-statistics.co/Linear-Regression.html}} represents the proportion of the variation in the measured real power explained by the model when taking the number of features in the model into consideration. A higher Adjusted R-Squared implies a better model. Mean Absolute Percentage Error (MAPE), defined as Equation \ref{eq:mape}, measures the prediction error, and a lower MAPE value indicates the better models' quality. Furthermore, Min/Max Accuracy is defined as Equation \ref{eq:minMax}, and it measures how far the model prediction is from the actual real power readings. The better the quality of the regression model is, the higher the Min/Max Accuracy will be. Finally, Correlation Accuracy is a simple correlation between predicted and actual real powers, and as the correlation among the values increases, it indicates that both values have similar directional moves. On average, the results report that power models on Ubuntu measure the power consumption with an accuracy of up to 95\% and in Windows with an accuracy of up to 85\%. We achieved better accuracy for Ubuntu because we had a shorter sampling rate, hence more accurate input. We also include a more detailed summary of all models in the appendix.

\begin{equation} \label{eq:minMax}
\begin{aligned}
MinMaxAccuracy = \\
mean\Bigg(\frac{min(actualRealPowers, predictedRealPowers)}{max(actualRealPowers, predictedRealPowers)}\Bigg)
\end{aligned}
\end{equation}

\begin{equation} \label{eq:mape}
\begin{aligned}
MAPE = \\ mean\Bigg(\frac{abs(predictedRealPowers - actualRealPowers)}{actualRealPowers}\Bigg)
\end{aligned}
\end{equation}

\begin{table}[h!] 
	\small
\caption{Power Model Evaluation Results}
\label{table:powerModelEvaluation}
\centering
\scalebox{0.8}{
\begin{tabular}{llccccc}
\hline \hline
\textit{OS} & \textit{Mode} & \textit{Adj $R^2$} & \textit{MAPE (\%)} & \textit{Min/Max ACC(\%)} & \textit{Correlation ACC}  \\ \hline
Ubuntu & Save & 0.968 & 5 & 95 & 0.984 \\
Ubuntu & Normal & 0.9065 & 8 & 93 & 0.954 \\
Win & Save & 0.8248 & 18 & 85 & 0.914 \\
Win & Normal & 0.5223 & 19 & 85 & 0.713 \\
\hline \hline
\end{tabular}
}
\end{table}

\begin{figure}[t!]
	\begin{center}
	\includegraphics[width=1\columnwidth]{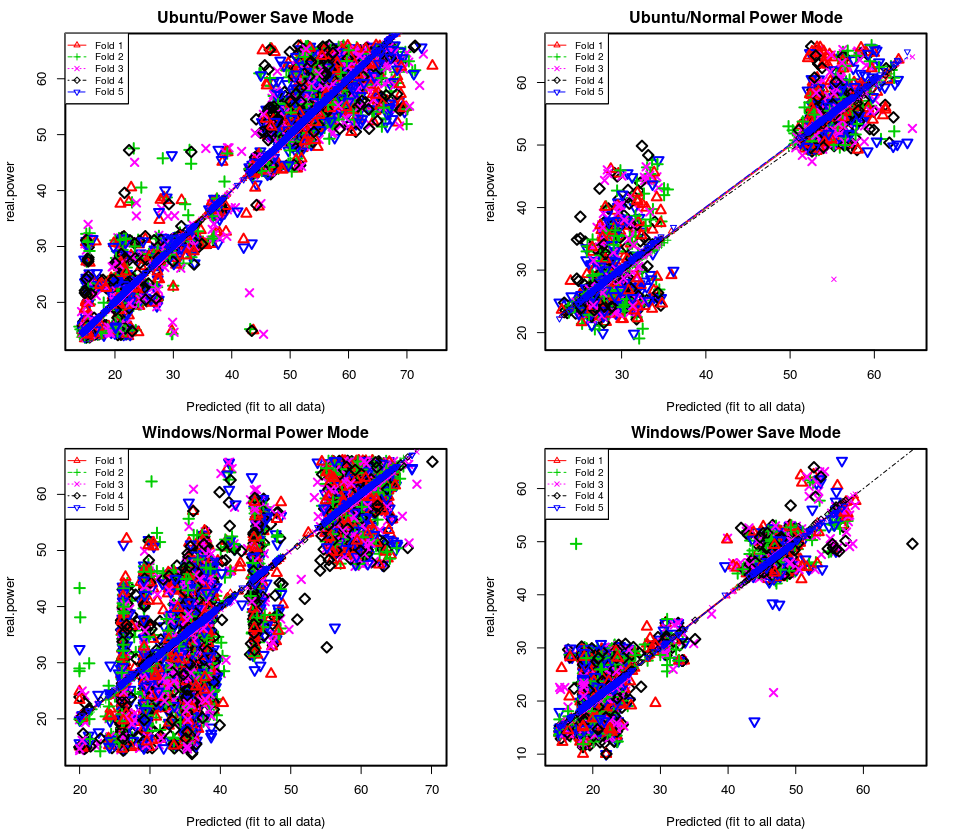}
	\end{center}
	\caption{Power models cross-validations.}
	\label{fig:modlCrossVal}
\end{figure}

Finally, we perform 5-folds cross-validations on all four models. Figure \ref{fig:modlCrossVal} illustrates all the resulting cross-validations side by side where the minor symbols are predicted real powers and bigger ones are actual real powers. Additionally, the mean squared error of the models varies from 8\% to 54\%, with power models of Ubuntu having a better performance than Windows models, as Table \ref{table:crossValResult} represents the average squared errors of five folds. According to the plots, we verify that model's prediction accuracies are approximately uniform among different samples, and the slopes and level of fitted lines have relatively low variations.

\begin{table}[h!] 
\caption{Power Models Cross-validations Mean Squared Error}
\label{table:crossValResult}
\centering
\begin{tabular}{llc}
\hline \hline
\textit{Model} & \textit{Power Mode} & \textit{Mean Squared Error} \\ \hline
Ubuntu&Save& 8.15 \\
Ubuntu&Normal& 33.44 \\
Windows&Save& 21.61 \\
Windows&Normal& 53.97 \\
\hline \hline
\end{tabular}
\end{table}

\subsection{i13DR Evaluation} 
\label{sec:i13drEvaluationOverview}

We designed a comprehensive experiment consisting of two parts to evaluate the performance of the i13DR infrastructure. The first part of the experiment is designed to estimate the i13DM capability to reduce the laptop's power consumption while connected to the electricity grid. For the second part of the experiment, we perform a demo scenario of a DR event with a few participating laptops to investigate the effect of the DR schedule on the demand load.

\subsubsection{Power Control Functionality Evaluation} 
\label{sec:i13drPowerControl} 

In our approach, we explained that our power consumption mechanism relies on the built-in functionalities of OS. To determine the impacts of i13DM's power control techniques on the laptop load curtailment, we executed various workloads on different OS with different power consumption modes. We used two workload generator applications to stress various components of the SUT. The SUT was fully charged to eliminate the impact of battery charging on the demand load. We executed a five-minute long workload on both OS; three times while the laptop was running in normal power consumption mode and three times while running on power save mode activated by i13DM. During the execution of workloads, we monitored and recorded the real power consumption in watts and consumed energy in kWh through a connected power meter. We used \textit{stress}\footnote{\url{http://manpages.ubuntu.com/manpages/zesty/man1/stress.1.html}} on Ubuntu to create workloads and on Windows, we made use of \textit{HeavyLoad v3.4}\footnote{\url{http://www.jam-software.com/heavyload}}.

We calculated the average power consumption drop in watts and energy consumption in kWh, as Table \ref{table:meanDrop} summarizes. The results indicate that i13DM can decrease the power consumption on Windows by up to 27\% and up to 7\% on Ubuntu. The main reason for such a significantly better performance is the availability of more extensive built-in power management features on Windows compared to Ubuntu. However, the drop in real power consumption on Windows is muchly noticeable than on Ubuntu. The reduction in total consumed energy is relatively in the same range for Windows and Ubuntu, with 27\% and 22\%, respectively. We believe the measurement device's low accuracy and low consumption of laptops over the short period of sampling are causing the similarity of observed energy consumption. For this reason, we argue that the drop in power consumption is a better indicator of i13DM's ability to control the electrical power consumption of laptops. 

\begin{table}[h!] 
\caption{Average Drop in Power and Energy Consumption}
\label{table:meanDrop}
\centering
\begin{tabular}{lccc}
\hline \hline
\textit{OS} & \textit{Power Drop (\%)} & \textit{Energy Drop (\%)}\\ \hline
Windows & 26.46 & 27.27 \\
Ubuntu & 6.95 & 22.22 \\
\hline \hline
\end{tabular}
\end{table}

\subsubsection{DR Event Scheduling Evaluation} 
\label{sec:i13DrEventEval} 

To investigate the effectiveness of i13DR in scheduling and performing DR events, we designed and conducted a field experiment for executing a DR event with a few participating laptops. We developed a scheduling mock-up component for i13DRP where the administrator can schedule and manage multiple DR events and monitor the participating laptops. Because i13DR was not connected to any RES supplies while performing the experiments, the mock-up component simulated the behavior of a wind turbine integrated with the local power grid. We also developed a scenario for scheduling a DR event initiated by an administrator. For planning a DR event, the administrator is first required to determine the position of the wind turbine. Afterward, the administrator specified the possible reduction of the wind turbine's electrical output in watts. The administrator was also required to provide the length of the DR event in minutes. The administrator also had the option to either start the DR event immediately or provide the start time of the DR event. Once all required inputs are provided, I13DRP starts the scheduling procedure. First, i13DRP queried all the currently online laptops to find the laptops within a 1000 meters radius of the provided location of the wind turbine. Next, for each retrieved laptop, i13DRP fetched the power consumption profiles from 20 minutes before the start time of the DR event up to the start time. Then, i13DRP accumulates the reported difference between real power consumption on average and power save modes from estimating the amount of power one specific laptop can contribute to power reduction. Finally, i13DRP created a schedule with the provided start time, or if the start time was not specified, i13DRP scheduled an event that immediately began and lasted as defined by the administrator. When all the schedules were created, i13DRP submitted the schedules to Firebase to be downloaded by i13DM on the filtered laptops. Afterward, i13DMs fetched the new schedule and activated and deactivated the power control according to the new timetable. Moreover, i13DMs sent a status code to Firebase announcing that they either joined or left the DR event, which the administrator could observe.

\subsubsection{Experimental Setup of the DR Event} 
\label{sec:i13DrEventSetupEval} 

We conducted the DR scheduling scenario on three laptops with the hardware specification in Table \ref{table:experimentDetails}. All the participating laptops were fully charged and connected to the MEDAL power meters, which recorded the power consumption during the experiments. We conducted the experiments five times which two DR events lasted for ten minutes and three events for five minutes. 

\begin{table}[h!] 
\small
\caption{Hardware Specification of Participating Laptops}
\label{table:experimentDetails}
\centering
\scalebox{0.80}{
\begin{tabular}{llllll}
\hline \hline
 \textit{Vendor} & \textit{OS} & \textit{CPU}  & \textit{RAM (GB)} & \textit{AC Output (w)}\\ \hline
 ThinkPad L540 & Win 10 & Intel Core i5 & 8   & 65  \\
 ThinkPad X230 & Ubuntu 16.04 & Intel Core i5 & 16  & 170  \\
 MacBook Pro & Win 10 & Intel Core i7 & 16  & 85 \\
\hline \hline
\end{tabular}
}
\end{table}

\subsubsection{Results of DR Events} 
\label{sec:i13DrEventResults} 

To evaluate the responsiveness of i13DR to the immediate changes in RES supply, we measured the latency of publishing a DR schedule to the real-time database after a request for a DR event was initiated. Furthermore, we measured the latency of all filtered laptops joining the DR event (fetching the DR schedule and activating the power control) after submitting the DR schedule to the real-time database by i13DRP. To evaluate the total impact of i13DR on curtailing the laptop demand load, we obtained the estimated demand reduction according to the laptops' power profiles and compared the estimated power reduction with the real power reduction provided from the readings of the MEDAL power meter. We also include the demand load plots for the participating laptops during the execution of each DR event in the appendix. 

On average, we observed a power reduction of 14.23 watts, although we expected an 8.92 watts reduction during the events, as Table \ref{table:summaryMeanPower2} illustrates. We also observed an average latency of  658 milliseconds for scheduling an event and an average latency of  164 milliseconds for all the selected laptops to join the DR event, as Table Table \ref{table:summaryDuraion} summarizes.

\begin{table}[h!] 
\caption{Summary of Experiments Average Power Reduction}
\label{table:summaryMeanPower2}
\centering
\scalebox{0.93}{
\begin{tabular}{cccccc}
\hline \hline
\textit{Index} & \textit{Est. Reduction (watts)}  & \textit{Average Reduction (watts)}  \\ \hline
1 & 9.48 & 9.75  \\
2 & 8.41 & 22.21  \\
3 & 9.75 & 15.35 \\
4 & 10.04 & 8.50  \\
5 & 6.93 & 15.34 \\ \hline
Average & 8.92 & 14.23 \\
\hline \hline
\end{tabular}}
\end{table}

\begin{table}[h!] 
\caption{Latency of DR Events}
\label{table:summaryDuraion}
\centering
\scalebox{.85}{
\begin{tabular}{cccccc}
\hline \hline
\textit{Index} & \textit{Scheduling Latency (ms)}  & \textit{Laptops Joining Latency (ms)} \\  \hline
1 & 654 & 103  \\
2 & 529 & 181  \\
3 & 726 & 217  \\
4 & 682 & 184  \\
5 & 698 & 135  \\ \hline 
Average & 658 & 164 \\
\hline \hline
\end{tabular}}
\end{table}	

Based on the results, we verify that i13DR can rapidly reduce the real power consumption of a group of laptops in a matter of milliseconds. In general, we observed a min/max accuracy of 66\% and a mean absolute percentage error of 72\% of the estimated demand reduction against the arithmetic mean of measured power consumption. However, the accuracy of our predictions is relatively low, which we discuss the reasons in the next section.

%% file: sections/discussion.tex
\section{Discussion and Limitations} 
\label{sec:disc}

Our findings show that the designed i13DR can successfully schedule and perform fast DR events. However, they also highlight some remarks and potential pitfalls. Although on average, the measured demand load reduction exceeds the estimated reduction, the accuracy of our estimation is relatively low at 66\% against the arithmetic mean of the measured power. The main reason for this relatively low accuracy is the low accuracy of deployed power models used for creating the power consumption profiles. All participating laptops used identical power models to estimate the power consumption, which is fitted based on the data collected from the SUT. Even though the power model yields relatively high accuracy for predicting real power consumption on SUT, the power model fails to measure the power on other laptops. Therefore, we argue that it is significantly vital to consider different hardware specifications when fitting the power consumption models. We ignore this factor in our design because we require additional measurement devices installed on the demand side, which are expensive for demand-side participants. Because one of our primary objectives is reducing the initial costs for participants, we sacrifice accuracy to reach our financial goals.

Since we used two Windows laptops and one Ubuntu laptop during the experiments, we expect to observe a 19.96\% reduction on average, according to previous evaluations. However, we observe a 17.80\% reduction according to the arithmetic mean of measured power consumption reductions, indicating a roughly 89\% accurate estimate of the effectiveness of power control approaches. 

During the experiments, we observed that i13DR could quickly respond to the immediate changes in RES. However,  the results also imply that i13DR significantly has higher latency for selecting the participating laptops and scheduling the DR events than the latency of all the laptops participating in the DR event. The primary reason for the more significant latency is the quality of the mock-up component developed for the experiments. The mock-up is created with AngularJS and is executed on the administrator's web browser. Therefore, the performance is limited to the administrator's laptop's performance (the SUT). Finally, we suggest developing the scheduling components as a server application that can offer to schedule a DR event as a service to the administrator. 

Finally, we notice that the amount of power reduction that laptops can contribute to the power grids using operating systems' built-in power management features is relatively low. Therefore, compensating for a substantial loss of RES supplies solely based on a limited number of participating laptops may not be feasible. 

%% file: sections/conclusions.tex
\section{Conclusions} 
\label{sec:conclusion}

In this study, we offer a design and an implementation of DR infrastructure for integrating the fluctuating into the grid. Furthermore, our approach offers no initial cost for demand-side participants and laptops, and our system can be used as a testbed for researching different scheduling and optimization approaches. The evaluation of our field experiments verifies that our system successfully schedules and executes DR events. Furthermore, we construct power models for estimating the power consumption of laptops with accuracy up to 95\% on Ubuntu and 85\% on Windows on the SUT. However, the accuracy of our estimation of demand load reduction is about 67\% indicating relatively low reliability. The main reason for the low accuracy is our entire software-based approach, which sacrifices high accuracy to eliminate initial DR costs. 

In future work, we suggest improving power modeling techniques, power consumption control methods, and DR event scheduling and optimization approaches. Our approach toward power modeling is solely based on linear regression techniques. However, there are several other machine learning approaches worth further studies, such as \textit{k-nearest neighbors algorithm}. Furthermore, we need to implement a production-ready scheduler and optimizer for minimizing the electrical load consumption on i13DRP, preferably based on the optimization algorithms such as \textit{Mixed Integer Linear Programming}, \textit{Convex Optimization Problem} and \textit{Particle Swarm Optimization}. Finally, as previously described, our power control approaches heavily depend on the built-in power management features of the OS. Although Windows offers extensive features in that regard, Ubuntu suffers from insufficient power management options. Therefore, we suggest investigating other techniques for managing power consumption on Ubuntu.

%% file: sections/appendix.tex
\appendix

\section{Appendix}

\subsection{Snapshots of i13DR}

\begin{figure}[h]
	\begin{center}
	\includegraphics[width=1\columnwidth]{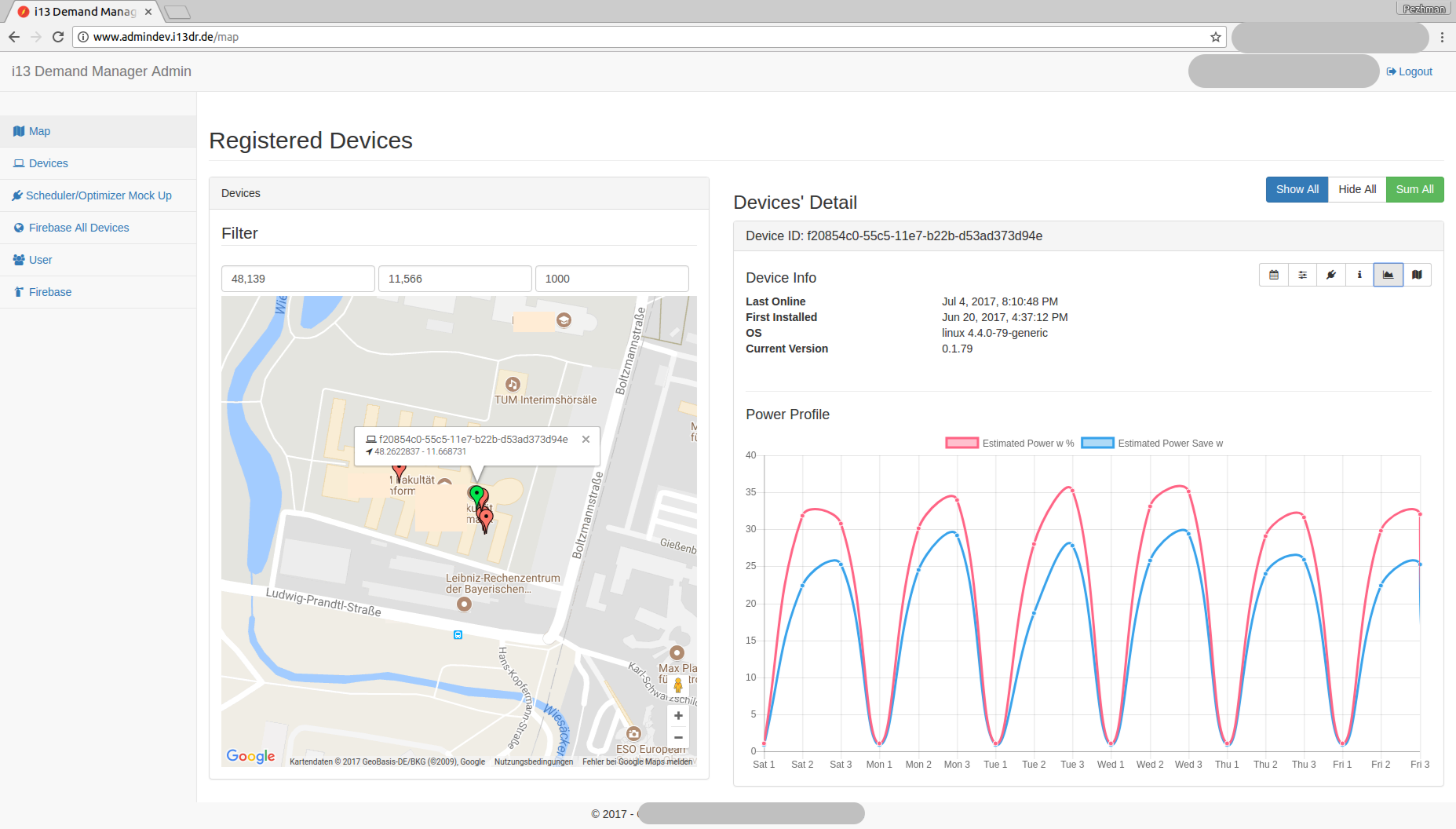}
	\end{center}
	\caption{A snapshot of i13DRP administrator panel with power profile of one laptop.}
	\label{fig:snp1}
\end{figure}

\begin{figure}[h]
	\begin{center}
	\includegraphics[width=1\columnwidth]{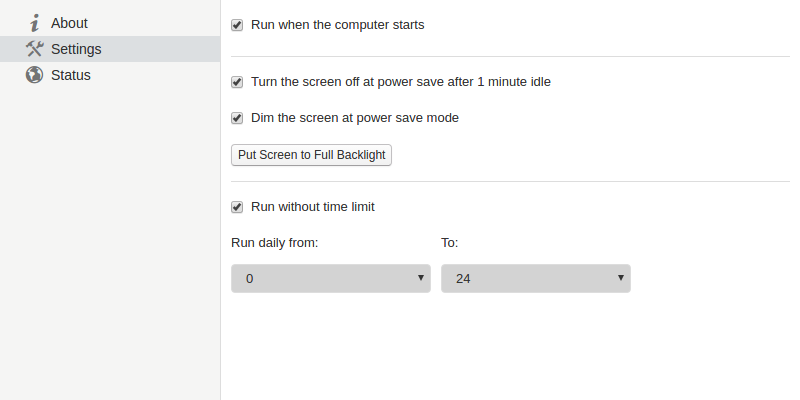}
	\end{center}
	\caption{A snapshot of i13DM's settings window on Ubuntu.}
	\label{fig:snp2}
\end{figure}

\begin{figure}[h]
	\begin{center}
	\includegraphics[width=1\columnwidth]{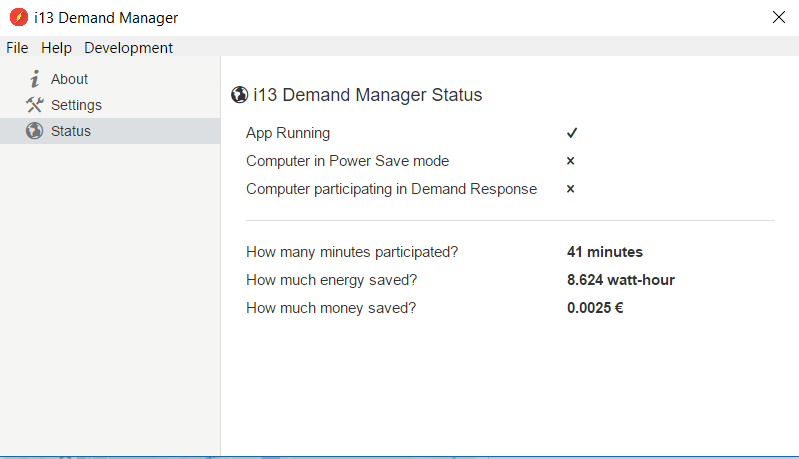}
	\end{center}
	\caption{A snapshot of i13DM's status window on Windows.}
	\label{fig:snp3}
\end{figure}

We include a few snapshots of i13DR where Figure \ref{fig:snp1} displays the i13DRP administrative page. Figure \ref{fig:snp2} and Figure \ref{fig:snp3} illustrate the i13DM's panel, where the user can see some reports on the laptop's power consumption and modify the behavior of i13DM.

\subsection{All Subsets Regressions of Power Related Data}

\begin{figure}[h]
   \centering
	\includegraphics[width=1\columnwidth]{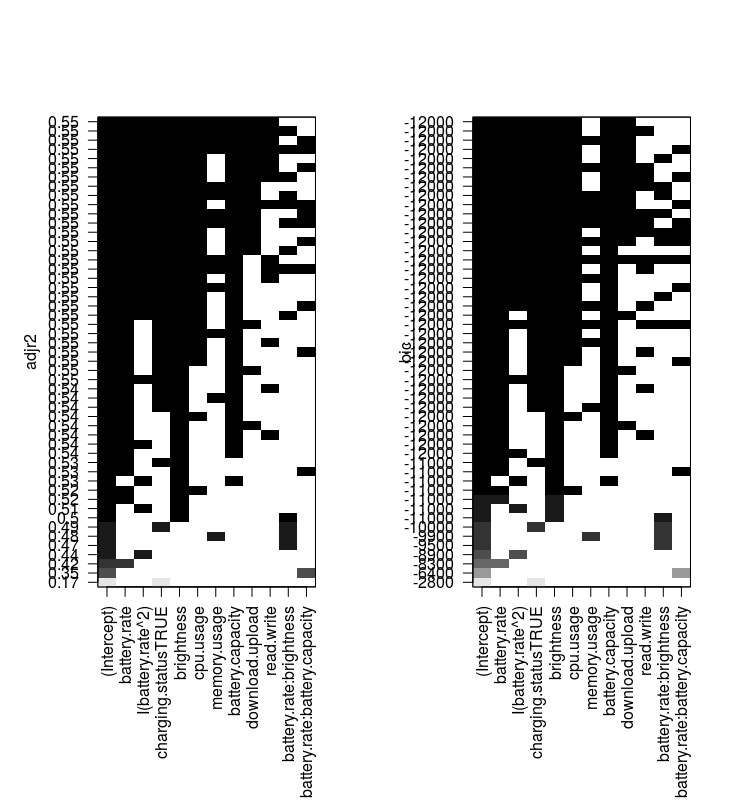}
	\caption{All subsets regressions for Windows in normal power mode.}
	\label{fig:leaps_w_n}
\end{figure}

\begin{figure}[h]
	\centering
	\includegraphics[width=1\columnwidth]{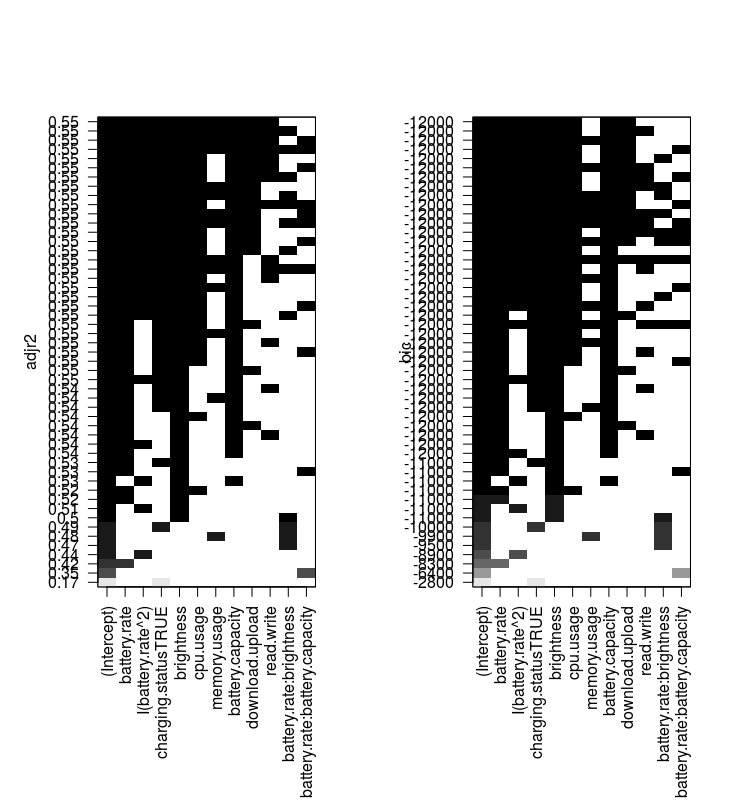}
	\caption{All subsets regressions for Windows in power save mode.}
	\label{fig:leaps_w_s}
\end{figure}

\begin{figure}[h]
	\centering
	\includegraphics[width=1\columnwidth]{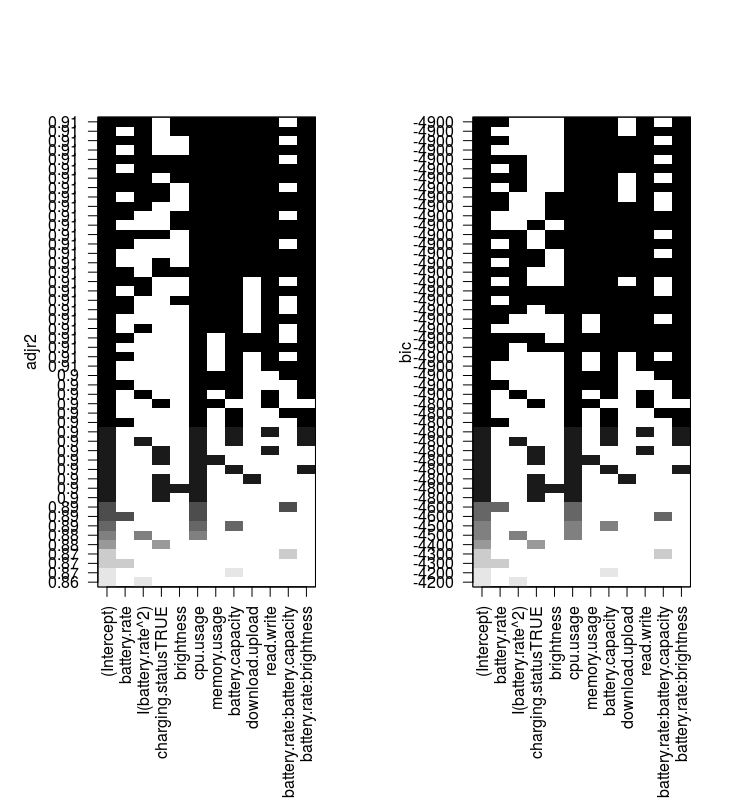}
	\caption{All subsets regressions for Ubuntu in normal power mode.}
	\label{fig:leaps_u_n}
\end{figure}

\begin{figure}[h]
	\centering
	\includegraphics[width=1\columnwidth]{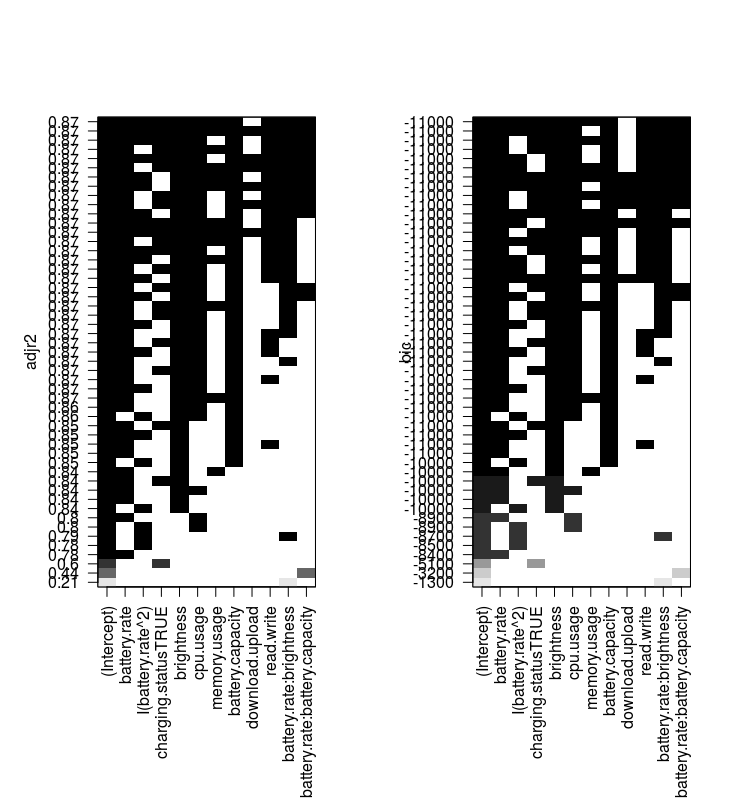}
	\caption{All subsets regressions for Ubuntu in power save mode.}
	\label{fig:leaps_u_s}
\end{figure}

We use \textit{R} package \textit{leaps}\footnote{\url{https://cran.r-project.org/web/packages/leaps/leaps.pdf}} to perform an exhaustive search to determine the best subset of power-related features. Figure \ref{fig:leaps_w_n}, Figure \ref{fig:leaps_w_s}, Figure \ref{fig:leaps_u_n}, and Figure \ref{fig:leaps_u_s} illustrates the result of an exhaustive search on all subsets of the regression model for the dataset collected from different OS in normal power consumption mode and power save mode. \textit{leaps} sorts the results by \textit{Adjusted R-Squared} and \textit{Bayesian Information Criterion (BIC)}\footnote{\url{http://r-statistics.co/Linear-Regression.html}} and examines them for the five best models reported for each subset size (one feature, two features and so on). The Adjusted R-Squared represents the proportion of variation in the measured power consumption that the model has explained by considering the number of features in the model. BIC measures the goodness of a model based on the maximized value of a likelihood function. The higher the calculated value for Adjusted R-squared, the better the constructed model, whereas as the value of BIC decreases, the model improves. 

\newpage

\newpage

\subsection{Power Models Summary}

\begin{figure}[h]
	\begin{center}
	\includegraphics[width=1\columnwidth]{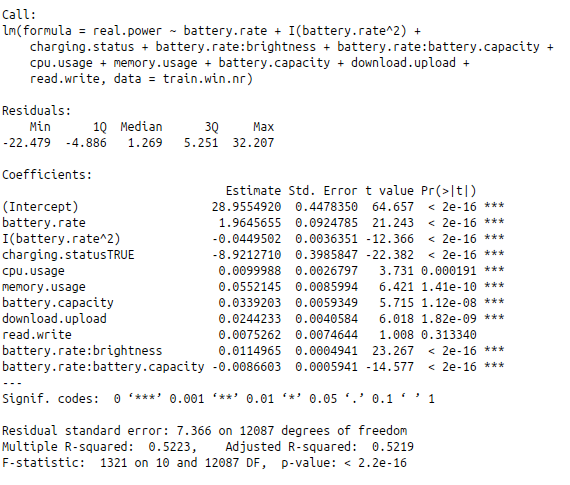}
	\end{center}
	\caption{Summary of power model for Windows on normal power mode.}
	\label{fig:sum1}
\end{figure}

\begin{figure}[h]
	\begin{center}
	\includegraphics[width=1\columnwidth]{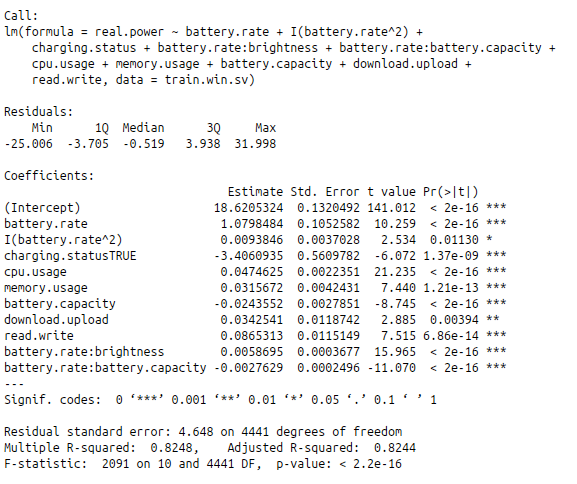}
	\end{center}
	\caption{Summary of power model for Windows on power save mode.}
	\label{fig:sum2}
\end{figure}

Figure \ref{fig:sum1}, Figure \ref{fig:sum2} and Figure \ref{fig:sum1} display the summary of each power model for different OS and power consumption mode, created by the \textit{summery()} method of \textit{R}.

\begin{figure}[h]
	\begin{center}
	\includegraphics[width=1\columnwidth]{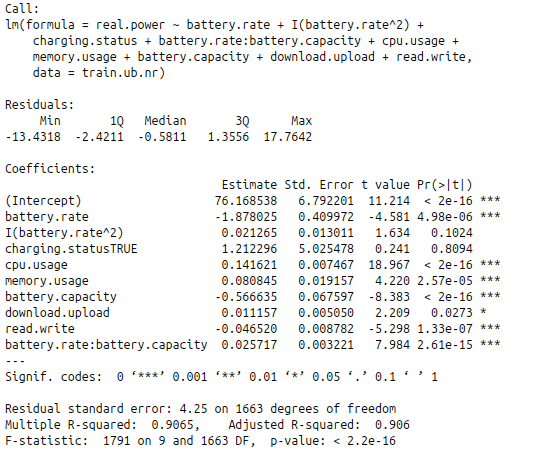}
	\end{center}
	\caption{Summary of power model for Ubuntu on normal power mode.}
	\label{fig:sum3}
\end{figure}

\begin{figure}[h]
	\begin{center}
	\includegraphics[width=1\columnwidth]{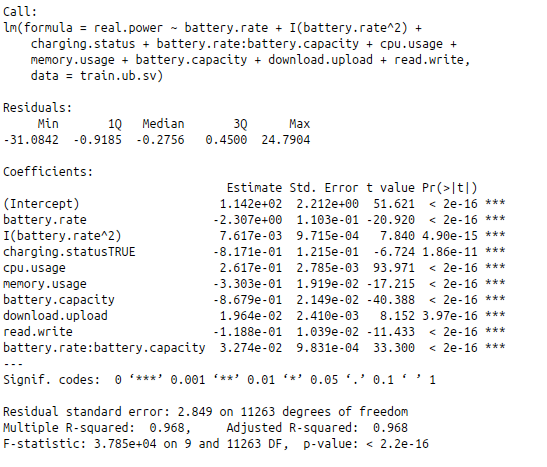}
	\end{center}
	\caption{Summary of power model for Ubuntu on power save mode.}
	\label{fig:sum4}
\end{figure}

\newpage

\subsection{Demand Load Plots of Experimental DR Events}

\begin{figure}[h]
	\begin{center}
	\includegraphics[width=1\columnwidth]{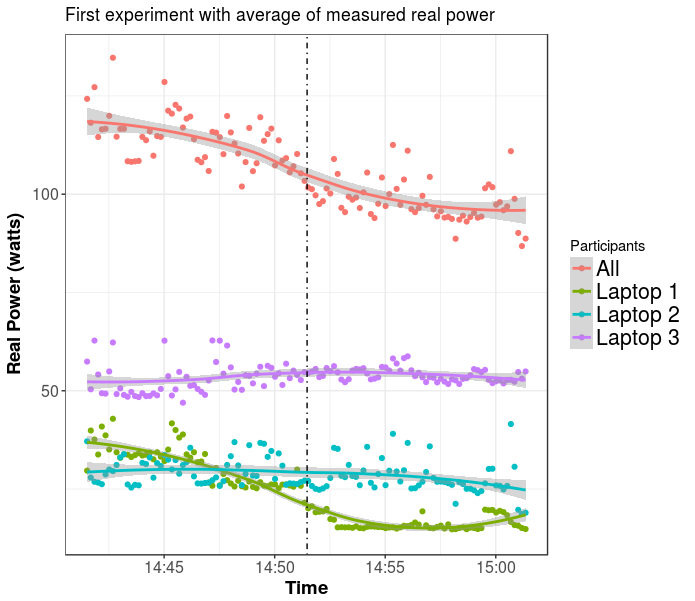}
	\end{center}
	\caption{First DR event.}
	\label{fig:event1}
\end{figure}

\begin{figure}[h]
	\begin{center}
	\includegraphics[width=1\columnwidth]{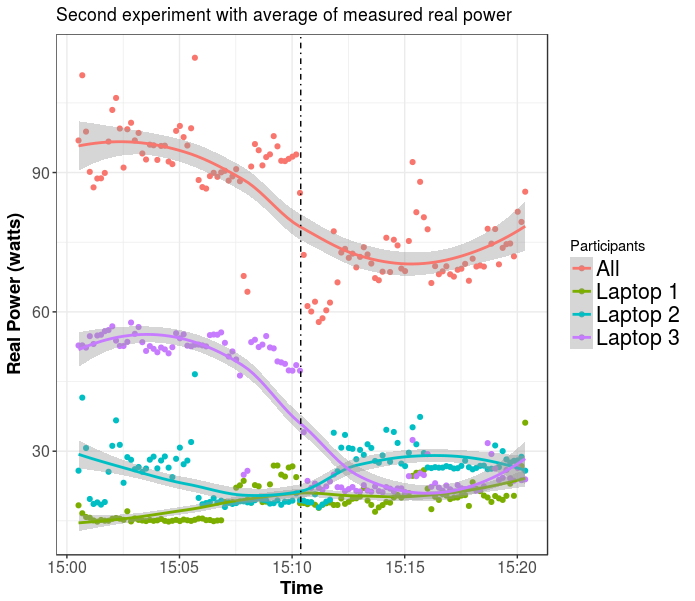}
	\end{center}
	\caption{Second DR event.}
	\label{fig:event2}
\end{figure}

\begin{figure}[h]
	\begin{center}
	\includegraphics[width=1\columnwidth]{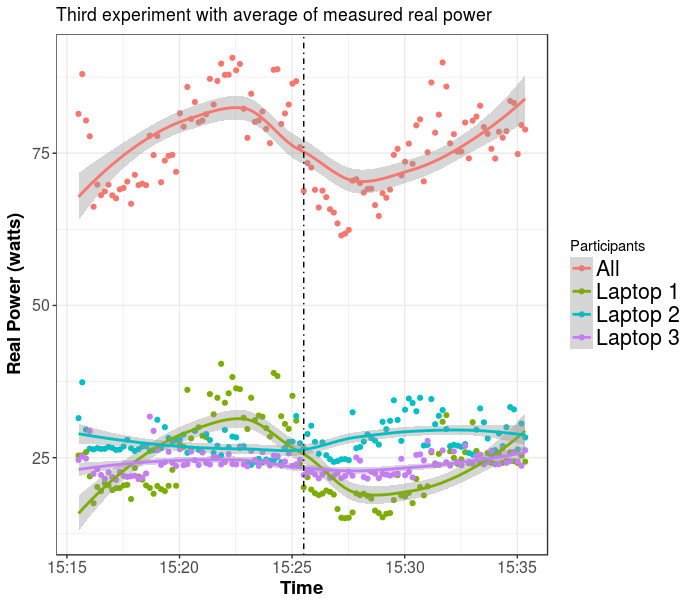}
	\end{center}
	\caption{Third DR event.}
	\label{fig:event3}
\end{figure}

\begin{figure}[h]
	\begin{center}
	\includegraphics[width=1\columnwidth]{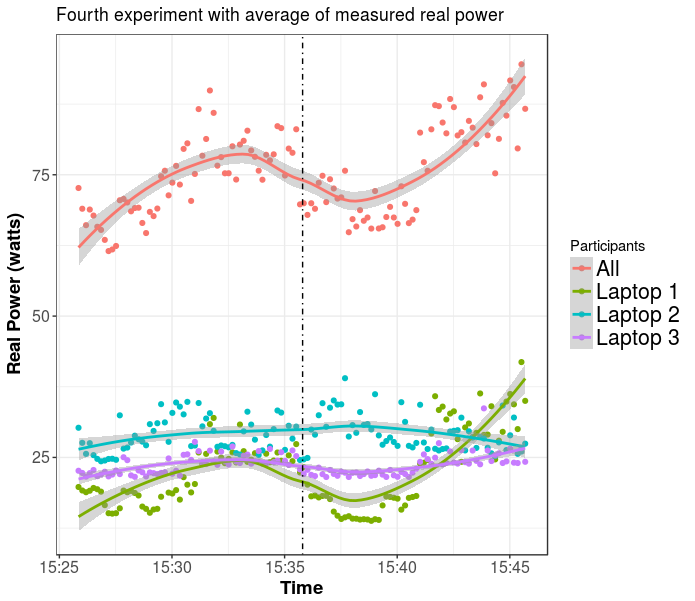}
	\end{center}
	\caption{Fourth DR event.}
	\label{fig:event4}
\end{figure}

\begin{figure}[h]
	\begin{center}
	\includegraphics[width=1\columnwidth]{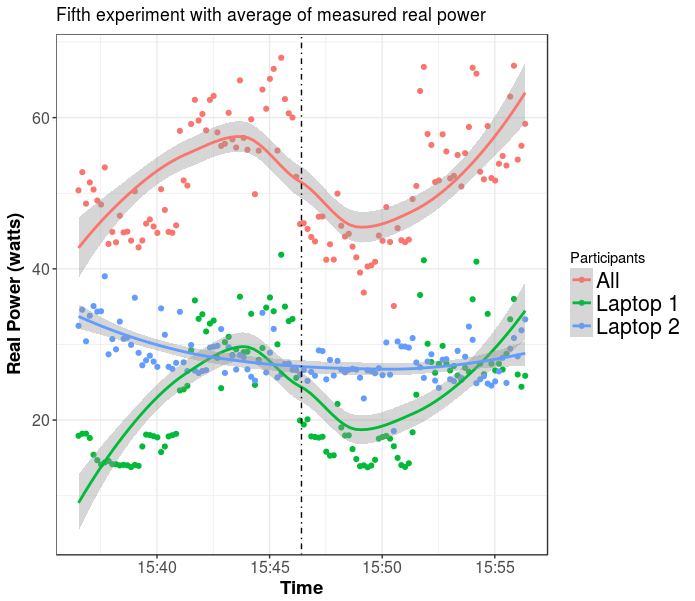}
	\end{center}
	\caption{Fifth DR event.}
	\label{fig:event5}
\end{figure}

Figure \ref{fig:event1}, Figure \ref{fig:event2}, Figure \ref{fig:event3}, Figure \ref{fig:event4}, and Figure \ref{fig:event5} illustrate the demand load plots for the participating laptops during the execution of each experiment, where the dotted line represent the moment power control is activated. Every plot includes the demand load of every laptop separately, named as \textit{Laptop 1}, \textit{Laptop 2} and \textit{Laptop 3} as well as the accumulated demand loads of all laptops, called as \textit{All}. Each plot demonstrates the demand load for ten minutes before and after the activation of power control independent of the length of the DR event.

